\documentstyle[12pt, epsf]{article}

\begin{document}
\baselineskip=0.7cm
\newcommand{\EQ}{\begin{equation}}
\newcommand{\EN}{\end{equation}}
\newcommand{\EQA}{\begin{eqnarray}}
\newcommand{\EQN}{\end{eqnarray}}
\newcommand{\EQAN}{\begin{eqnarray*}}
\newcommand{\EQNN}{\end{eqnarray*}}
\newcommand{\e}{{\rm e}}
\newcommand{\Sp}{{\rm Sp}}
\renewcommand{\theequation}{\arabic{section}.\arabic{equation}}
\newcommand{\Tr}{{\rm Tr}}
\newcommand{\lpartial}{\buildrel \leftarrow \over \partial}
\newcommand{\rpartial}{\buildrel \rightarrow \over 
\partial}
\renewcommand{\thesection}{\arabic{section}.}
\renewcommand{\thesubsection}{\arabic{section}.\arabic{subsection}}
\makeatletter
\def\section{\@startsection{section}{1}{\z@}{-3.5ex plus -1ex minus 
 -.2ex}{2.3ex plus .2ex}{\large}} 
\def\subsection{\@startsection{subsection}{2}{\z@}{-3.25ex plus -1ex minus 
 -.2ex}{1.5ex plus .2ex}{\normalsize\it}}
\makeatother
\def\thefootnote{\fnsymbol{footnote}}

\begin{flushright}
hep-th/0210243\\
UT-KOMABA/02-13\\
October 2002
\end{flushright}
\vspace{1cm}
\begin{center}
\Large

From Wrapped Supermembrane to 
M(atrix) Theory
\footnote{Presented at the 3rd Sakharov International Conference on Physics, 
Moscow, June, 2002.}

\vspace{1cm}
\normalsize
 Tamiaki {\sc  Yoneya}
\footnote{
e-mail address:\ \ {\tt tam@hep1.c.u-tokyo.ac.jp}}
\\
\vspace{0.3cm}

{\it Institute of Physics, University of Tokyo\\
Komaba, Meguro-ku, Tokyo 153-8902}

\vspace{0.6cm}
Abstract
\end{center}
The issue of justifying the matrix-theory proposal is revisited. 
We first discuss  how the matrix-string theory is derived directly 
starting from the eleven dimensional
supermembrane wrapped  around a circle of radius $R=g_s\ell_s$, 
without invoking any stringy  assumptions, such as S- and
T-dualities.  This derivation provides us a basis for studying 
both string ($R\rightarrow 0)$- and M ($R\rightarrow \infty$)-theory 
 limits of quantum membrane theory in a single unified 
framework. 
In particular, we show that two different boosts 
of supermembrane, namely 
one of unwrapped membrane along the M-theory circle 
and the other of membrane wrapped 
 about a transervse direction which is orthogonal 
to the M-theory circle, give the same matrix theory 
in the 11 dimensional limit, $R\rightarrow \infty$ 
(with $N\rightarrow \infty$). 
We also discuss briefly the nature of possible 
covariantized matrix (string) theories. 

\newpage
\section{Introduction}
It has often been stated that string theory is no more 
a theory of strings, since 
we have a plentiful set of objects of varied dimensionalities. 
However, fundamental strings still play a privileged role, 
in the sense that the dynamics of these objects should  be formulated as various collective degrees of 
freedom in the theory. 
As far as we know, only legitimate descriptions of the dynamics of 
such branes are through strings, unless we are satisfied with 
some effective low-energy approximations as substitutes for 
exact description in principle. 
However, there have been a couple of proposals for exact gauge-theory 
models, 
which go beyond mere effective descriptions of branes and 
point towards possibilities of exact nonperturbative 
theories. One (and actually the first) 
of the most notable examples of such proposals 
has been the so-called M(atrix) theory advocated in \cite{bfss}, as 
a possible exact formulation of M-theory in a particular 
infinite momentum frame (IMF), in which the system is 
infinitely boosted along the 11-th compactified 
direction of M-theory.  
Unfortunately, however, it seems fair to assess that 
no substantial  progress has been made in 
recent few years on M(atrix) theory, after initial explosion of 
papers until around 1998. 

In this note, I would like to revisit Matrix theory
 from a viewpoint of 
its connection with supermembrane. 
Just as the fundamental strings play a special role in 
the ordinary perturbative string theory, it is expected that 
supermembrane would play a pivotal role 
in the quest for the fundamental degrees of freedom in M-theory. 
In fact, it is well known that M(atrix) theory can be interpreted 
as a particular (regularized) realization of quantum 
supermembrane theory by discretizing  membrane degrees of freedom
in the form of matrices.  On the other hand, in the usual
interpretation, the fundamental strings of type IIA string theory
should be elevated to supermembranes, which are  wrapped along
the 11-th direction.  Combining these viewpoints, we expect that
IIA superstrings  should also be directly related to M(atrix)
theory.   It is somewhat surprising, however, that these two very 
familiar viewpoints, IMF and wrapped 
supermembrane, have  never been formulated in 
a really coherent and unified framework. 
Here I would like to fill this gap with a hope that 
such a work might shed new light on this important question 
and would stimulate further investigations along this line. 

We organize the present note as follows. 
First in section 2 we briefly 
recapitulate the original proposals of Matrix theory and 
Matrix-string theory. In section 3, we present a new derivation 
of Matrix-string theory \cite{dvv} 
without relying on  string duality arguments,  as
a new direct regularization of  wrapped supermembrane,
following a previous work
\cite{sekinoyoneya}.  Then, in section 4, we discuss the
decompactification limit of  our formulation of Matrix-string
theory  and show that it leads to the same Matrix-theory action 
as the one we have from the original BFSS proposal. 
This shows that two apparent different ways 
of matrix regularizations with different directions 
of boosting to light velocity lead to the same theory 
in the decompactification limit of M-theory. 
It seems that these observations shed a useful new 
light on the interpretation of both string and membrane. 
In the final section, we conclude with considerations 
on the nature of covariantized matrix (string) theories.

\section{Revisiting Matrix-theory proposal}

Let us begin by recalling the BFSS proposal. 
There were two key backgrounds for this proposal:
On the one hand, it has been well known that
 the low-energy effective action
for the dynamics of D-particles is the maximally supersymmetric 
quantum mechanics with 9-independent matrix coordinates 
$X^a$ $(a=1,2, \ldots, 9)$, which can be 
obtained by dimensional reduction 
from the maximally supersymmetric 
Yang-Mills theory in 9+1 dimensional space-time. 
On the other hand, essentially the same model 
can be derived as a regularized version of 
supermembrane in 11 dimensions in the light-cone 
gauge, as has been known from the work \cite{dhn}. 
The action density takes the forms, depending on these two 
interpretations, respectively, as 
\EQ
\Tr 
\Bigl( {1\over 2g_s \ell_s} D_{\tau} X^a D_{\tau} X^a + i \theta^T
D_{\tau}
\theta  +{1 \over \ell_s^5 } [X^a, X^b]^2 -
{1\over \ell_s^2}\theta^T \gamma_a [\theta, X^a]\bigr) , 
\EN
or
\EQ
\Tr 
\Bigl( {1\over 2R} D_{\tau} X^a D_{\tau} X^a + i \theta^T D_{\tau}
\theta  +{R \over 4\ell_P^6 } [X^a, X^b]^2 -
{R\over \ell_P^3}\theta^T \gamma_a [\theta, X^a]\bigr),  
\EN
where the first one is written using the standard 
string coupling $g_s$, in terms of which the 
mass of D-particle is $1/g_s\ell_s$, with string 
length constant $\ell_s$,
while,   in the second,  using 11 dimensional radius $R=g_s\ell_s$
and  11 dimensional Planck length $\ell_P=g_s^{1/3}\ell_s$. 

The low-energy approximation is justified when 
the massive modes of the fundamental open strings 
attached to D-particles are decoupled, that is to say, 
when the typical energy scale $E$ in D-particle 
dynamics is much lower than 
the excitation energy $\sim 1/\ell_s$ of strings. 
The typical energy of open strings connecting 
D-particles is roughly of the order of the distances 
among D-particles when the velocity is sufficiently small. 
Hence, we can justify the low-energy description 
if the string coupling is  sufficiently small 
and the distances among D-particles are small compared with 
the string length, 
\EQ
r\ll \ell_s, \quad g_s\ll 1 , 
\label{lowcond}
\EN
which is consistently implemented since the 
characteristic spatial length scale of the system 
is equal to the 11-dimensional Planck length $\ell_P$.

On the other hand, the same theory written in the second form 
as a regularization \cite{dhn} of
supermembrane  could be a good constructive definition of 
the quantum supermembrane theory when the size of matrices
$N\times N$ is  sufficiently large,  
$
N\rightarrow \infty.
$

The proposal of Matrix theory can be regarded as the 
unification of these two viewpoints on the same action. 
Supposing that the (total) momentum 
of the system of $N$ D-particles along the 11-th 
direction is sufficiently large compared with other 
directions, we can identify it with the longitudinal 
momentum of the light-cone frame as
\EQ
{N\over R}= P^{11}\sim P^+=(P^0+P^{11})/2. 
\label{imfcond}
\EN
Indeed, for $P^{11}>0$, 
the Hamiltonian of this system can be 
written in the form 
\EQ
H(=-P^-)={(P^a)^2\over 2P^+} + \cdots. 
\EN
where $P^a$ are the transverse momentum corresponding to 
the center of mass coordinate of $N$ D-particles. 
Anti-D-particles corresponding to $P^{11}<0$ 
are decoupled in this limit. 
For finite $N$, $g_s$ must be sufficiently small 
for the validity of (\ref{imfcond}). This is consistent with 
(\ref{lowcond})  \cite{seibergsen}. 
This means that  the second 
interpretation as the light-cone description 
of 11 dimensional M-theory may be meaningful even for
finite $N$,  irrespectively of the 
membrane picture, in the limit of small string coupling. 
Furthermore, we can keep the validity of 
the relation (\ref{imfcond})  when 
$g_s$ becomes large if we assume sufficiently large $N$. 
This suggests that the constructive definition of
supermembrane  theory in the light-like frame 
could  in turn be interpreted as the
theory of large number ($=N$) of D-particles. 

Obviously, this is only suggestive and never shows anything 
which are really required to justify that the theory defines an 
{\it exact} description of D-particle dynamics. A crucial question is 
whether the process of taking large $N$ limit recovers  
the whole stringy degrees (or their appropriate 
extension for large $g_s$) of freedom, which
must be  essential for the emergence of 
consistent quantum theory of gravity. 
On dimensional grounds, we can expect, remembering  the 
relation $\ell_P=g_s^{1/3}\ell_s$, that 
the string excitations might somehow be eliminated 
since the string scale can be taken to be 
infinitely small compared with the 11D Planck length 
in the limit of large string coupling. However, 
in order to utilize this feature, 
we need some concrete theoretical framework which 
is defined in 11 dimensions as an extension of string theory. 
Such a theory must exhibit correct stringy degrees of 
freedom in the 10D limit and is desirably to be 
consistent with 11 dimensional supergravity 
in the low-energy 11D limit. 
But it would almost amount to constructing M-theory!

Thus, instead of directly approaching 
this question, we can ask from a slightly different angle 
the following question: How, if any, is the above picture 
related to another familiar picture that 
the fundamental string should be a wrapped supermembrane, 
and how is it related with the above matrix model? 
In ref. \cite{sekinoyoneya}, we have discussed this question by 
studying a natural regularization for wrapped supermembrane. 
We here briefly summarize this result. 

\section{Matrix string from wrapped supermembrane}

We start from the action of supermembrane 
in the light-cone gauge \cite{dhn}. 
\[
A = {1\over \ell_P^3}\int d\tau \int_0^{2\pi L}d\sigma \int_0^{2\pi
L} d\rho \, \, {\cal L}, 
\]
\EQ
w^{-1}{\cal L}
={1\over 2}(D_0 X^a)^2 +i \overline{\psi}\gamma_-D_0 \psi
-{1\over 4}(\{X^a, X^b\})^2 
+i\overline{\psi}\gamma_-\Gamma_a\{X^a, \psi \} ,
\label{lightconemembrane}
\EN
with
\[
\{X^a, X^b\}=w^{-1}(\partial_{\sigma}X^a\partial_{\rho}
X^b -\partial_{\rho}X^a\partial_{\sigma} 
X^b)\, \quad D_0 X^a= \partial_{\tau} X^a-\{A_0, X^a\} .
\]
The density function $w$ is introduced in the 
gauge fixing process such that the 
longitudinal momentum $P^+(\sigma, \rho)$ 
satisfies $P^+(\sigma, \rho)=P^+w(\sigma, \rho)/L^2$. 
The time coordinate 
$\tau$ is related to the light-cone time by 
\EQ
\ell_M^3P^+\tau/(2\pi L)^2=X^+, 
\label{timerelation}
\EN
 such 
that the total center of mass (transverse) momentum $P^a$ 
contributes to the Hamiltonian 
in the standard form, 
\[
P^-dX^+ =\Big({(P^a)^2\over 2P^+} + \cdots\Big)dX^+.
\] 
The length parameter $L$ is arbitrary of which the theory is 
independent, as can be checked easily by performing suitable 
rescaling. 
It is well known that this action can be regarded as a
$(0+1)$-dimensional gauge  theory whose gauge group is the
group of area-preserving  diffeomorphisms (APD) of
two-dimensional  space $(\sigma, \rho)$. In fact, 
the Gauss-law constraint derived
from   this action by the variation 
with respect to the gauge field $A_0$ gives the 
constraint corresponding to the 
area-preserving diffeomorphism (APD) 
which is the residual reparametrization symmetry
$\delta X^a =\{\Lambda, X\}, \, \, \delta 
A_0 =\partial_0 \Lambda+\{\Lambda, A_0\}, \, etc$, 
 after 
fixing to the light-cone gauge. More precisely, 
the Gauss-law constraint, 
\EQ
\{D_0 X^a, X^a\} +i\{\overline{\psi}, \gamma_-\psi\}=0 ,
\label{gausslaw}
\EN
is the integrability condition for the equation 
determining the longitudinal coordinate $X^-$.  

We can now assume that one, say 9-th,  of the 9 transverse
directions  ($a, b, \ldots, =1, 2,\ldots, 9$) to be 
the compactified M-theory direction of radius $R=
g_s\ell_s$. It is easy to check that 
if we simply make the `double-dimensional reduction' 
by ignoring the dependence on the 
world-volume coordinate along the wrapped direction 
(which we choose to be $\rho$), the action 
just reduces to the standard Green-Schwarz action of 
IIA theory in the light-cone gauge. 
This is consistent with the 
familiar statement that IIA string is identified 
with wrapped supermembrane in the limit of 
small compactification radius. 
However, for any finite $g_s$, it is not justifiable 
to neglect the $\rho$-dependence. 
The reason is that there is {\it no mas gap} which would 
characterize the compactification on the world volume. 
Remember that the usual kinetic term 
of the form $(\partial_{\rho}X)^2$ does not 
exist in the action. 
Therefore it is absolutely necessary in quantum theory 
to examine 
the form of the action without making the naive 
reduction. 

Let us call the 9-th transverse coordinate $Y$. 
Wrapping can be 
explicitly taken into account by making a shift 
of $Y$ as 
\EQ
Y\rightarrow \rho + Y ,
\EN
and assuming that the redefined $Y$ as well as other
components  and fermion coordinates are periodic in $\rho$. 
By performing the rescaling $(\sigma, \tau, \rho) \rightarrow 
R(\sigma, \tau, \rho)$ and choosing $L$ to be 
identified with $R$, the above shift 
amounts to making the following replacements in the 
action
\[
{1\over 2}(D_0 Y)^2\rightarrow 
R^{-2}{1\over 2}(\partial_0 Y-
\partial_{\sigma}A -{1\over R}
\{A, Y\})^2 , 
\]
\[
{1\over 2}\{X^i, Y\}^2 \rightarrow R^{-2}{1\over 2}
(\partial_{\sigma} X^i -{1\over R}\{Y, X^i\})^2 ,
\]
\[
\psi^T\Gamma_9\{Y, \psi\}
\rightarrow -R^{-1}\psi^T\Gamma_9(\partial_{\sigma}\psi 
-{1\over R}\{Y, \psi\}).
\]
Then the action now takes {\it almost} (but not quite)
  the form of
an 
$(1+1)$-dimensional  gauge theory with gauge group APD, 
\[
A={1\over \ell_s^2}
\int d\tau \int_0^{2\pi}
d\sigma
\int_0^{2\pi}d\rho 
\Big[
{1\over 2}F_{0, \sigma}^2+
{1\over 2}(D_0X^i)^2 -{1\over 2}(D^Y_{\sigma}X^i)^2
\]
\EQ\hspace{5cm}
-{1\over 4R^2}\{X^i, X^j\}^2
+ \cdots \Big] ,
\label{wrappedmemaction}
\EN
in the sense that we now have the covariant derivative 
for both $\tau$ and $\sigma$ directions and 
the field strength $F_{0,\sigma}$. The peculiarity 
preventing to be the gauge theory in its strict sense 
is that the would-be internal space and base space 
overlap along the $\sigma$-direction. Here the transverse 
indices $(i, j, \ldots, )$ run over only 8 directions, 
since the 9-th direction $Y$ turned into the gauge field 
of $\sigma$-direction. 

What we have found in \cite{sekinoyoneya}\footnote{
For further works, appeared after the conference, 
discussing  this construction from somewhat different angles, 
see  
\cite{hi}, \cite{ceder}. 
} is that 
there is a natural regularization which converts the above 
action into that of Matrix-string theory \cite{dvv}. 
The rule for making correspondence between 
matrix string and membrane is exemplified by the following 
table
\vspace{0.5cm}
\begin{center}
\begin{tabular}{c|c}
 Long string of matrix string theory &  
Doubly compactified membrane \\ \hline\hline
& \\

$\Tr {1\over N}\int^{2\pi}_0 d\theta $ & 
$\int_0^{2\pi} d\sigma {1\over 2\pi}\int_0^{2\pi} d\rho$ \\ 

& \\
$i{N\over 2\pi}[A, B]$ & $\{A, B \}$ \\

& \\ \hline

& \\

$A_{k\ell}(\theta)$ & 
$\int {d\rho\over 2\pi}  e^{-in\rho} A(\sigma, \rho)  $ \\

& \\
$k-\ell$ & $n$ \\

&\\
$\theta$  & $\sigma =
{(k+\ell-2)\pi \over N}+{\theta\over N}$
\\ & \\
\hline
\end{tabular}
\end{center}
\vspace{0.5cm}
Thus the basic idea is the identification of 
the nonzero Fourier modes along the compactified direction 
$\rho$ with the off-diagonal matrix elements of $(1+1)$-dimensional 
matrix field $A_{k\ell}(\theta, \tau)$ and, 
simultaneously, the matrix elements with fixed 
$k-\ell$ with $N$ bits of strings. 
For more detail, see our original paper. 
Based on this idea, we 
can establish the general formula connecting 
matrix-string and membrane:
\EQA
&&{1\over N}\int d\theta \, \Tr\big(
M^{(1)}(\theta)M^{(2)}(\theta) \cdots 
M^{(\ell)}(\theta)\big)\nonumber 
\\
&&\hspace{-1.2cm}
={1\over 2\pi}
\int d\rho\int d\sigma 
\exp \Big[ -i{\pi\over N}\sum_{\ell\ge i>j\ge 1}
(\partial_{\sigma_j}\partial_{\rho_i}
-\partial_{\rho_j}\partial_{\sigma_i}) \Big]\\
&&\hspace{3cm}
\times \, \, M^{(1)}(\sigma_1, \rho_1)\cdots  
M^{(\ell)}(\sigma_{\ell},\rho_{\ell})
\Big|_{\sigma_i =\sigma, \rho_i=\rho} .
\nonumber
\EQN 
As a special case of this formula, we have, say, 
\[
{1\over N}\int d\theta \, {\rm STr}
\Big([M^{(1)}(\theta), M^{(2)}(\theta)]
[M^{(3)}(\theta), M^{(4)}(\theta)]M^{(3)}(\theta)
\cdots M^{(\ell)}(\theta)\Big)
\]
\[
=
-{1\over 2\pi}(2\pi/N)^2 
\int d\sigma d\rho\, 
\{M^{(1)}(\sigma,\rho), M^{(2)}(\sigma, \rho)\}
\{M^{(3)}(\sigma, \rho), M^{(4)}(\sigma, \rho)\}
\]
\EQ
\hspace{0.8cm}\times  M^{(3)}(\sigma, \rho)
\cdots M^{(\ell)}(\sigma, \rho)
(1+O(1/N^2))
\label{geneformula2}
\EN 
in the large $N$ limit, 
where the symmetrized trace (STr) means to 
treat the commutators on the left hand side 
as single matrices. 

These prescription allows us to regularize the 
membrane action (\ref{wrappedmemaction}) in the 
matrix form as,  
up to $O(1/N^2)$ corrections, 
\EQA
A={(2\pi)^2L\over \ell_P^3}&&\hspace{-0.7cm}
\int d\tau {2\pi \over N}\int_0^{2\pi}
d\theta \, 
\Tr\Big(
{1\over 2}F_{0, \theta}^2+
{1\over 2}(D_0X^i)^2 -{1\over 2}N^2(D_{\theta}X^i)^2
\nonumber \\
&&\hspace{-3cm}
+{1\over 4L^2}({N\over 2\pi})^2[X^i, X^j]^2
+i\psi^T D_0\psi -Ni\, \psi^T\Gamma_9D_{\theta}\psi
-{1\over L}{N\over 2\pi}\psi^T\Gamma_i[X^i,\psi]
\Big) ,
\label{mataction}
\EQN
where 
\EQ
D_{\theta}X^i =\partial_{\theta}X^i-i{1\over 2\pi L}[Y, X^i] ,
\EN
\EQ
D_0 X^i =\partial_{\tau}X^i-i{N\over 2\pi L}[A, X^i] ,
\EN
\EQ
F_{0,\theta}
=\partial_{\tau}Y -N\partial_{\theta}A
-i{N\over 2\pi L}[A, Y] ,
\EN 
and similarly for fermion variables. 
By performing the redefinition 
\[
\tau \rightarrow \tau/N, \quad
L\rightarrow L/2\pi, \quad \psi \rightarrow \sqrt{N}\psi,  
\] 
the $N$ dependence is eliminated and the action is 
reduced to  the standard matrix-string theory action.   
Recall $L=R$ and $R/\ell_P^3=1/\ell_s^2$. 
Assuming that the physical light-cone time 
$X^+$ must be independent on $N$, 
this rescaling of the time coordinate requires, 
by the relation (\ref{timerelation}),  that 
the total longitudinal momentum $P^+$ scales 
with $N$ as 
\EQ
P^+ \rightarrow NP^+
\label{pplus}
\EN
which coincides with the correct scaling for the 
matrix-string theory interpretation. 

The merit of this new derivation of matrix-string 
theory is that we have not assumed 
any string dualities (both S and T dualities) 
which have been invoked as a justification of this model 
in the original proposal
\cite{dvv}. The usual method \cite{taylor} of 
performing T-duality for 
matrix models is based on the low-energy 
approximation where 
higher string modes are not taken into account 
explicitly. Also, 
S-duality still of course remains as an unproven 
conjecture. Thus, due to our new formulation, the matrix-string
theory  can now be regarded as a natural regularization 
of supermembrane theory on an equal footing
as  Matrix theory, independently of the 
string duality arguments. Comparing with the 
matrix theory, the matrix-string action has a more 
direct connection with the ordinary perturbative 
formulation of IIA strings, in such a way that 
D-particles (and possibly higher D-branes) 
can also be clearly identified as various 
conserved charges. In particular, the D-particle charge, 
being the momentum along the compactified 9-th direction, 
is related to the electric flux.  

\section{Matrix theory from matrix string}
We now come to 
the question how IIA string, as wrapped supermembrane,  
is related to Matrix theory which has been obtained 
as another possible regularization of supermembrane 
boosted along the compactified M-theory direction 
without wrapping.    

Before going to that, let us briefly recall 
how the matrix-string action (\ref{mataction}) reduces to 
IIA string theory in the 10D limit $g_s\rightarrow 0$. 
Since, (\ref{mataction}) being 
the action of an (1+1)-dimensional
gauge  theory, the coupling constant is inversely 
proportional to $R=g_s\ell_s$, the 10D limit is 
an IR limit (or strong coupling limit) of the gauge theory. 
If we fix the string scale $\ell_s$, 11 dimensional Planck 
scale becomes negligible in the limit. 
It is then reasonable to expect that 
the quantum fluctuations are restricted to the so-called  
flat directions where the potential vanishes, 
\EQ
[X^i, X^j]\Rightarrow
 0 \quad \mbox{for all transverse directions
$(i, j)$}. 
\EN
This is the orbifold ($\sim ({\bf R}^8)^N/S_N$) 2D CFT \cite{dvv}
which has been argued to be  equivalent, in the large $N$ limit,
with the Green-Schwarz  formulation of IIA  string theory. As the
above correspondence  between matrix-string action and the wrapped
membrane  shows, this reduction to diagonal matrices is 
just equivalent to the simple  KK reduction ignoring higher Fourier
modes of membrane excitations along the wrapped direction. 
Therefore rigorous justification of this reduction for 
arbitrarily small but nonzero string coupling should be 
very nontrivial \cite{sekinoyoneya}, owing to the absence
of mass gap  associated with this reduction or wrapping. 

Now we consider the opposite limit $R\rightarrow \infty$ 
(strong string coupling), which 
is, in the language of 2D gauge theory, the weak or UV limit. 
From the viewpoint of M-theory, we have to fix the M-theory 
scale $\ell_P=g_s^{1/3}\ell_s$. Thus the string scale now 
becomes infinite $\ell_s=g_s^{-1/3}\ell_P\rightarrow 0$. 
Since in our formulation the matrix-string action is derived
entirely  within the  logic of 11 dimensions and  is connected,
with the proviso as being warned above, smoothly to  string
theory, it can provide a concrete 
{\it intermediate} framework which we
argued  to be necessary for discussing
the  M-theory limit of matrix models.  
That is still an intermediate step in the sense that 
we have not yet any guarantee for the compatibility of matrix-string action 
with 11D supergravity. 

Performing the rescaling $\tau \rightarrow (\ell_P/\ell_s^2)\tau$, 
the action is rewritten as 
\EQ
\int d\tau\int_0^{2\pi}
d\theta \, 
{\ell_P\over \ell_s^4}\Tr\Big(
{1\over 2}F_{0, \theta}^2+
{1\over 2}(D_0X^i)^2 -{1\over 2}(D_{\theta}X^i)^2
+{\ell_s^4\over 4\ell_P^6}[X^i, X^j]^2
+\cdots\Big)
\EN 
with 
\EQ
F_{0, \theta}={\ell_s^2\over \ell_P}\partial_0A_{\theta}
-\partial_{\theta}A_0-{i\ell_s^2\over \ell_P^3}
[A_{\theta}, A_0], 
\EN
\EQ
D_0X^i={\ell_s^2\over  \ell_P}\partial_0 -{i\ell_s^2\over
\ell_P^3} [A_0, X^i], 
\EN
\EQ
D_{\theta}X^i=\partial_{\theta}X^i-{i\ell_s^2\over \ell_P^3}
[A_{\theta}, X^i], \quad etc.
\EN
Let us adopt the axial gauge $\partial_{\theta}A_{\theta}=0$. 
Now, from the above form of the action, 
we can see easily that, in the
limit
$\ell_s\rightarrow 0$ with $\ell_P$ kept fixed,  
the quantum fluctuations  are restricted to
those satisfying 
\EQ
\partial_{\theta}X^i
=\partial_{\theta}A_0=\partial_{\theta}\psi=0. 
\EN
Together with our gauge condition, this shows that in the M-theory 
limit we can ignore $\theta$-dependence of the matrix variables.  
Thus the effective action 
for the remaining fluctuations is given by
\EQ
{2\pi\over \ell_P}\int d\tau \, 
\Tr\Big(
{1\over 2}(D_0X^a)^2  
+{1\over 4\ell_P^4}[X^a, X^b]^2
+\cdots\Big), 
\label{uvlimitaction}
\EN
where now 
\[
D_0X^a=\partial_0X^a-{i\over \ell_P^2}[A_0, X^a], \quad etc, 
\]
with the  indices $a, b, \ldots$ running over 9 coordinates,
including 
$A_{\theta}=X^9$. Of course, the large $N$-limit must 
be understood for this action, since the model can 
be justified only in the large $N$-limit in the 
10D limit. Intuitively, the situation is as follows: As 
wrapped membrane is stretched infinitely along the wrapped
direction,  its size with respect to 10 dimensional world 
is infinitely squeezed and the resulting thin tube along the 
wrapped direction is torn off into an infinite number of 
`bits' of gravitons represented by the matrices $X_a$.

The above result (\ref{uvlimitaction}) obviously takes the 
similar form as that of Matrix theory. In fact, 
they are identical if one performs the rescaling 
$\tau\rightarrow (\ell_P/R)\tau, X^a\rightarrow X^a$ to the 
standard form of the Matrix-theory action. The necessity of 
this rescaling is related to the fact that the 
typical temporal and spatial scales characterizing the
Matrix-theory  action are $\tau_c\sim g_s^{-2/3}\ell_P$ and 
$\ell_P$, respectively. This conforms to the stringy
uncertainty relation, 
$\Delta T \Delta X \ge \ell_s^2=\ell_P^3/R$,  
of space-time \cite{yospacetime}. The rescaling
$\tau\rightarrow (\ell_P/R)\tau$ is just  appropriate 
for converting the temporal scale to the Planck scale. 

\section{Towards covariantized matrix theories}
Now what should we  learn from this new derivation of 
Matrix-theory? 
First,  in the case of wrapped 
supermembrane, the 9-th transverse direction  is identified
with  the compactified M-theory direction before taking the 
11D limit $R\rightarrow \infty$. On the other hand, 
in the original BFSS proposal, 
the M-theory direction is the longitudinal direction along 
which the system is boosted to IMF. Thus in the latter case, 
the transverse directions are completely orthogonal to 
the M-theory direction. Yet, we found that in the
decompactification limit, both give the same theory. 
This may be an indication that  Matrix theory 
in the decompactification limit is Lorentz covariant 
as one hopes.
 
Second, as we have already emphasized, our 11D derivation of
matrix-string theory  can play the role of a unified framework 
for two different but mutually complementary pictures, 
wrapped supermembrane and IMF along the compactified direction, 
on the  relation between 10D and 11D physics. 
From this viewpoint, a very important next step 
seems to construct the 
covariantized version of matrix-string
theory.  In the rest of this note, let us briefly 
compare the possible directions of covariantized 
matrix models. 

We call the direct covariantization of Matrix theory 
possibility `pIMF', and the indirect covariantization 
through wrapped membrane picture possibility `pWM', 
respectively. 
In pIMF, we require the existence of some generalized matrix
variables representing the dynamics of D-particles in general 
Lorentz frame, such that they reduce, under the infinite boost 
along the compactified direction, to the usual matrices of the
standard light-cone  formulation. Such generalized matrix
variables  must necessarily represent an infinite number of both
D-particles and anti-D-particles 
simultaneously, and the  generic elements of the
generalized matrix variables  must correspond to massless modes 
of open strings connecting D-and/or anti-D-particles. 
However, for the open strings connecting between 
D-and anti-D-particles, supersymmetry is spontaneously broken. 
If we include only the massless modes of strings, the 
representation of spontaneously broken IIA supersymmetry 
must be nonlinear, since the number of massless modes 
are different between space-time bosons and fermions for those 
strings  corresponding to 
a `wrong' GSO projection. Therefore,
in the  approach pIMF, the possibility of constructing
covariant  Matrix theory would crucially  depend on obtaining 
an appropriate concrete nonlinear realization of IIA
supersymmetry.  From the viewpoint of 
continuum membrane theory, it seems natural to adopt 
a supermembrane action {\it without} $\kappa$-symmetry. 
The action is required to  reduce to the standard action 
constructed by requiring $\kappa$-symmetry in the light-cone gauge, 
when the system is boosted to IMF. Thus some of the
(generalized) matrix  degrees of freedom must be decoupled
dynamically in the IMF limit. One of the puzzles in pursuing 
this possibility is that Lorentz transformation would 
in general involve the change of the physical degrees of 
freedom. If we boost the system in the positive direction 
along the M-theory direction, some ($\Delta N$) of the anti-D-particles 
turn into D-particles. From the viewpoint of the effective 
low-energy theory, this implies that the effective gauge group 
changes as SU($N_+$)$\times$SU($N_-$) $\rightarrow$ SU($N_+ +\Delta N$)
$\times$SU($N_- -\Delta N$), which leads to the corresponding 
change with respect to the numbers of 
gauge fields and massless fermions. It seems very nontrivial to 
achieve this in any manifestly Lorentz covariant theory 
for large but finite $N$. 
Hopefully, this may be a hint in looking for the 
correct ``generalized" matrix variables. 

In the approach pWM, on the other hand, 
the situation is very different. In this case, 
both D-particles and anti-D-particles can be 
taken into account by considering states with positive and 
negative momenta along the 9-th direction. 
The spontaneously broken susy would be 
realized in the presence of whole massive string modes 
by the emergence of world sheet fermion fields with 
non-standard boundary conditions for fermionic 
Green-Schwarz (GS) open  strings
connected to  D-and anti-D-particles, as discussed in
\cite{yosusy} in detail. 
  From the viewpoint of supermembrane, the action must 
be $\kappa$-symmetric, since in the 10D limit it should 
reduce to the ordinary GS IIA string. 
The broken susy must automatically appear in the 
sectors with coexisting D-and anti-D-particles in the 
above sense. In the 11D limit with appropriate large $N$
limit,  the theory must be  equivalent with what one arrives
in the approach pIMF.  We  do not see any fundamental
conceptual problem,  in contrast to the 
approach pIFM, 
in applying our new method of regularizing wrapped 
membranes in
the  possible covariant formulations, though there appear 
some technical subtleties in realizing the (anti)
D-particles in terms of the picture of wrapped membrane. 
A promising framework along this direction 
seems to be the recent interesting proposal \cite{berkovits} 
by Berkovits using the pure-spinor formalism. 
Once a reasonable covariantized matrix-string theory 
could be constructed, the next question would be whether 
it gives 11D supergravity correctly in the long-distance
limit.  Assuming that its 10D limit describes the correct
perturbative IIA string theory, it seems very plausible that 
the correct covariant 11D extension is nothing but 11D
supergravity,  although no general theorem of this sort is
known.  We hope to return to this challenging question in the
near future.

\vspace{0.3cm}
\noindent
Acknowledgements: 
I would like to thank the organizers of the 3rd Sakharov
International  Conference on Physics for inviting me. 
The present work is partially supported by Grant-in-Aid 
for Scientific Research (No. 12440060 and No. 13135205) 
from the Ministry of Education, Science and Culture. 


\end{document}